\documentstyle[prl,aps,epsf]{revtex}

\begin{document}
\def \beq {\begin{equation}}
\def \eeq {\end{equation}}
\def \bes {\begin{eqnarray}}
\def \ees {\end{eqnarray}}
\def \ni {\noindent}
\def \nn {\nonumber}
\def \z {\tilde{z}}
\def \kp {k_{\bot}}
\def \e {\varepsilon}
\def \dpa {\Delta_{\|}}
\def \dpp {\Delta_{\bot}}

\title{Correlation of energy and free energy for the 
thermal Casimir force between real metals}

\author{
V.~B.~Bezerra,
G.~L.~Klimchitskaya
and V.~M.~Mostepanenko
}
\address
{Departamento de F\'{\i}sica, Universidade Federal da Para\'{\i}ba,
C.P.~5008, CEP 58059-970,
Jo\~{a}o Pessoa, Pb-Brazil
}
\maketitle

{\abstract{The energy of fluctuating electromagnetic field is 
investigated for the thermal Casimir force acting between
parallel plates made of real metal. It is proved that for nondissipative
media with temperature independent dielectric permittivity the energy
at nonzero temperature comprises of the (renormalized) energies of
the zero-point and thermal photons. In this manner photons can be
considered as collective elementary excitations of the matter of plates and
electromagnetic field. If the dielectric permittivity depends on
temperature the energy contains additional terms proportional to the 
derivatives of $\e$ with respect to temperature, and the quasiparticle
interpretation of the fluctuating field is not possible. The correlation
between energy and free energy is considered. Previous calculations 
of the Casimir energy in the 
framework of the Lifshitz formula at zero temperature and optical
tabulated data supplemented by the Drude model at room temperature
are analysed. It is demonstrated that this quantity is not 
a good approximation either for the free energy or the energy. A physical
interpretation of this hybrid quantity is suggested. The contradictory
results in the recent literature on whether the zero-frequency
term of the Lifshitz formula for the perpendicular polarized modes
has any effective contribution to 
the physical quantities are discussed. Four main approaches to
the resolution of this problem are specified. The precise expressions for 
entropy of the fluctuating field between plates made of real metal are
obtained, which helps to decide between the different approaches. 
The conclusion
is that the Lifshitz formula supplemented by the plasma model and
the surface impedance approach are best suited to describe the thermal 
Casimir force between real metals.
}}


\section{Introduction}

The Casimir effect is a rare macroscopic manifestation of the zero-point
electromagnetic energy. It results from the alteration of the zero-point
spectrum by the material boundaries (see the original Casimir's paper
\cite{1} and extensive reviews \cite{2,3,4,5,6}). In recent years 
considerable attention was paid to the precision measurements of the
Casimir force between metallic surfaces \cite{7,8,9,10,11,12,13,14,15}.
The results of these measurements were used in Refs.~\cite{16,17,18,19,20}
to constrain the hypothetical interactions predicted by many extensions to 
the Standard Model and also in nanotechnology \cite{21}. This called for
new theoretical investigation of the Casimir force with allowance made
for the realistic boundary properties, i.e., surface roughness, finite
conductivity of a metal and nonzero temperature (see Ref.~\cite{6} for
the re\-view).
Also the combined effect of these factors has attracted considerable attention.

In this paper we consider the correlation between energy and free energy for
the Casimir force acting between two plane parallel plates made of real
metal of finite conductivity kept at nonzero temperature. At zero
temperature the influence of the finite conductivity of the boundary metal
onto the Casimir force was examined in Refs.~\cite{22,23,23a,24} on the basis
of the Lifshitz theory using the optical tabulated data supplemented by
the Drude model. In this approach the Casimir energy was represented by an
integral with respect to continuous frequency, as it is at zero temperature.
However, the optical tabulated data and the value of the relaxation parameter
of the Drude model at $T=300\,$K were substituted, which are actually
temperature dependent. By this means, in fact, some hybrid quantity was
computed, which is 
different from both energy and free energy, with no
clear relevance to either of them. When it is considered that the Casimir
energy between two plates is generally used to calculate the Casimir force
in the experimental configuration of a sphere (spherical lens) above a plate,
the resolution of this issue is of great interest. Below is shown the
relationship between energy, free energy and a quantity computed in
Refs.~\cite{22,23,23a,24}.

Investigation of the correlation between energy and free energy 
for the thermal 
Casimir force has also assumed great importance in connection with the
contradictory results obtained by different authors when applying the
Lifshitz theory to real metals \cite{25,26,27,39,28,29,30,31,32,33}.
The contradictions arise on whether or not the zero-frequency term of the
Lifshitz formula for the perpendicular polarized modes of electromagnetic
field contribute to physical quantities. There are four main approaches
to the resolution of this issue in the recent literature.

a) According to Refs.~\cite{25,33} based on the immediate application of
the unmodified Lifshitz formula, the zero-frequency term of this formula for
the perpendicular polarized modes is equal to zero in the case of real
metals described by the Drude model (remind 
the reader that for ideal metals the
reflection coefficients for both polarizations are equal to unity at zero
frequency, and hence the zero-frequency term for both modes is not equal
to zero). This approach leads to the conclusion that
thermal corrections to the Casimir force  are large,
negative, and linear in temperature
at small separations, and the asymptotic Casimir force between real
metals at large separations is two times smaller than for the case of
an ideal metal (with no regard for the particular value of the
conductivity) \cite{30}.

b) From the standpoint of Refs.~\cite{27,30}, the perpendicular polarized 
modes give a nonzero contribution at zero frequency. To find it, a special 
modification of the zero-frequency term of the Lifshitz formula was
proposed (found by analogy with the prescription of Ref.~\cite{34} for an 
ideal metal but not coinciding with it).

c) According to Refs.~\cite{28,29}, the perpendicular polarized modes 
also give 
a nonzero contribution at zero frequency. For real metals it is, however,
the same as for ideal metal, i.e., the reflection coefficients for both
modes are equal to unity at zero frequency. What this means is for real
metals the same modification of the zero-frequency term of the Lifshitz
formula is made as for ideal metals \cite{34}. This approach leads to 
linear (although positive) thermal corrections to the Casimir force at 
small separations and to the absence of any finite conductivity
corrections for real metals starting from moderate separations of several
micrometers regardless of metal quality \cite{30}.

d) Finally, according to the approach of Refs.~\cite{26,27,39} both modes
with the parallel and perpendicular polarizations do contribute to the
zero-frequency term and this contribution can
be calculated by the substitution
of the plasma model dielectric function into the unmodified Lifshitz formula.
The same conclusion is obtained in Ref.~\cite{31} on the basis of the
surface impedance approach.

Thus, at the moment there is no agreement 
in the theoretical literature as to
the description of the thermal Casimir force between real metals.
To gain a more complete understanding of the present state of affairs,
in Ref.~\cite{32} the thermodynamical argument was exploited. According
to Ref.~\cite{32}, the approaches a) and c) do not conform to the
requirements of thermodynamics as they lead to the negative values of
entropy and violation of the Nernst heat theorem. Although the qualitative
conclusions of \cite{32} are quite correct, the quantitative calculations
are incomplete as they do not take into account the entropy of real photons.
The precise expressions for the energy and free energy of the 
fluctuating electromagnetic field 
found below are used to obtain the quantitative behavior of entropy
as a function of surface separation distance and temperature. The obtained
results confirm the conclusion of Ref.~\cite{32} that the approaches a) 
and c) are not compatible with thermodynamics. They also give the 
possibility to compare the approaches b) and d) in order to decide
between them.

The paper is organized as follows. In Sec.~II the main notations are 
introduced and the case of nondissipative condensed  media 
separated by a gap is considered,
with the media described by a temperature independent
dielectric permittivity. It is proved that in this situation one can
introduce photons as quasiparticles due to the collective elementary
excitations of condensed matter and electromagnetic field. As a consequence,
the energy at temperature $T$ defined via the derivative of the free energy
with respect to $T$ comprises of the (renormalized) energies of the zero-point
and thermal photons. In Sec.~III the energy and free energy of the  
fluctuating electromagnetic field 
are considered on the basis of the Lifshitz theory and 
the plasma model. In Sec.~IV the applicability of the Drude model for the
calculation of the thermal Casimir force is discussed. The energy and free
energy are found in the case of a metal described by the Drude model.
The alternative approaches to this problem available in literature are
analysed and compared. The physical sense of the energy at temperature $T$
appears to be more complicated than in the case of the plasma model. It
is shown that energy contains additional terms depending on the derivatives of
the dielectric permittivity with respect to temperature. In Sec.~V the 
entropy for the thermal Casimir force acting between real metals is
calculated precisely for both plasma and Drude dielectric functions.
Sec.~VI contains conclusions and discussion. 

\section{Photons between plates as elementary excitations}

We consider the configuration of two semispaces (thick plates)
with frequency-dependent dielectric permittivity ${\e}(\omega)$
restricted by parallel planes and separated by an empty space
with distance $a$ between them at a temperature $T$. This is a
system in thermal equilibrium. The free energy per unit area is
given by the well known Lifshitz formula \cite{2,6,34a,35}
\beq
F_E(a,T)=\frac{k_B T}{4\pi}
\sum_{l=-\infty}^{\infty}
\int_{0}^{\infty}
{\kp}d{\kp}\left[ \ln{\dpa}(\xi_l,{\kp})
+
\ln{\dpp}(\xi_l,{\kp})\right].
\label{1}
\eeq
\ni
Here $\Delta_{\|,\bot}(\omega,\kp)$ are the quantities having zero
values on the photon eigenfrequencies permitted between the plates
by the boundary conditions (indices $\|,\,\bot$ label two independent
polarizations, and $\kp$ is the modulus of a wave vector in the
plane of plates)
\beq
\Delta_{\|,\bot}(\omega_{\kp,n}^{\|,\bot},\kp)=0.
\label{2}
\eeq
\ni
They can be expressed in terms of reflection coefficients
on the imaginary frequency axis
\beq
\Delta_{\|,\bot}(\xi_l,\kp)=1-r_{\|,\bot}^2(\xi_l,\kp)e^{-2aq_l},
\label{3}
\eeq
\ni
where
\beq
r_{\|}^2(\xi_l,\kp)=\left[\frac{{\e}(i\xi)q_l-k_l}{{\e}(i\xi)q_l+k_l}
\right]^2\!\!\!\!, {\quad }
r_{\bot}^2(\xi_l,\kp)=\left(\frac{q_l-k_l}{q_l+k_l}\right)^2
\label{4}
\eeq
\ni
with the notations
\beq
q_l\equiv\sqrt{\frac{\xi_l^2}{c^2}+k_{\bot}^2},
\qquad
k_l\equiv\sqrt{{\e}(i\xi_l)\frac{\xi_l^2}{c^2}+k_{\bot}^2}.
\label{5}
\eeq
\ni
In Eq.(\ref{1}) $k_B$ is the Boltzmann constant and
$\xi_l=2\pi lk_B T/\hbar$, where $l=0,\pm1,\pm2,\ldots\,$, are
the Matsubara frequencies. As seen from Eq.(\ref{3}), the quantities
$\Delta_{\|,\bot}$ are normalized in such a way that the free energy
(\ref{1}) tends to zero for the infinitely remote plates 
($a\to\infty$). The details of the renormalization procedure can be
found in Refs.\cite{6,24}.

In this section we consider nondissipative media, which is
to say that ${\e}(\omega)$ is a real function. It will be assumed also that
at a given frequency $\e$ does not depend on temperature. Both conditions
are satisfied, e.g., for metals described by the plasma model or for
dielectrics with a constant dielectric permittivity (the case of dissipative
media is considered in Secs.IV,V). Under these conditions we find the simple
quasiparticle interpretation for photons between plates and the expression
for the energy of the fluctuating field at a temperature $T$.

According to thermodynamics, energy at arbitrary temperature is given by
\beq
E(a,T)=-T^2\frac{\partial}{\partial T}\frac{F_E(a,T)}{T},
\label{6}
\eeq
\ni
where the free energy is defined in Eq.(\ref{1}).

Taking into account that the term of Eq.(\ref{1}) with $l=0$ is linear
in temperature and the quantities $\Delta_{\|,\bot}$ are even functions of 
$l$ one obtains
\beq
E(a,T)=-\frac{k_B T^2}{2\pi}
\sum_{l=1}^{\infty}
\int_{0}^{\infty}
{\kp}d{\kp}\frac{\partial}{\partial T}\left[ \ln{\dpa}(\xi_l,{\kp})
+\ln{\dpp}(\xi_l,{\kp})\right].
\label{7}
\eeq
\ni
Let us next use that $\Delta_{\|,\bot}$ depend on temperature
through the Matsubara frequencies only, so that
$\partial/\partial T=(\xi_l/T)\partial/\partial \xi_l$. Thus, 
energy per unit area is given by
\beq
E(a,T)=-\frac{k_B T}{2\pi}
\sum_{l=1}^{\infty}\xi_l
\int_{0}^{\infty}
{\kp}d{\kp}\frac{\partial}{\partial \xi_l}\left[ \ln{\dpa}(\xi_l,{\kp})
+\ln{\dpp}(\xi_l,{\kp})\right].
\label{8}
\eeq

We consider now the interpretation of energy at temperature $T$ in terms 
of elementary excitations. In the case of the nondissipative media under
consideration the photon eigenfrequencies are real and the nonrenormalized
energy of equilibrium fluctuating electromagnetic field in the system
comprises of the energy of zero-point fluctuations and Planck's protons
\cite{36}
\beq
E_{nr}(a,T)=\hbar
\int_{0}^{\infty}
\frac{{\kp}d{\kp}}{2\pi}
\sum_{n}\left\{\omega_{{\kp},n}^{\|}\left[\frac{1}{2}+
\frac{1}{e^{\hbar \omega_{{\kp},n}^{\|}/(k_B T)}-1}\right]
+\omega_{{\kp},n}^{\bot}\left[\frac{1}{2}+
\frac{1}{e^{\hbar \omega_{{\kp},n}^{\bot}/(k_B T)}-1}\right]\right\}.
\label{9}
\eeq

Identically, Eq.(\ref{9}) can be rearranged to give
\beq
E_{nr}(a,T)=\frac{\hbar}{2}
\int_{0}^{\infty}
\frac{{\kp}d{\kp}}{2\pi}
\sum_{n}\left(\omega_{{\kp},n}^{\|}\coth
\frac{\hbar \omega_{{\kp},n}^{\|}}{2k_B T}
+
\omega_{{\kp},n}^{\bot}\coth
\frac{\hbar \omega_{{\kp},n}^{\bot}}{2k_B T}\right).
\label{10}
\eeq

This quantity is evidently infinite. The renormalized value of the sum
over the eigenfrequencies $\omega_{{\kp},n}^{\|,\bot}$ can be calculated
by the use of the argument theorem \cite{2,3,6,24,37}.
\beq
\sum_{n}\left(\omega_{{\kp},n}^{\|}\coth
\frac{\hbar \omega_{{\kp},n}^{\|}}{2k_B T}
+
\omega_{{\kp},n}^{\bot}\coth
\frac{\hbar \omega_{{\kp},n}^{\bot}}{2k_B T}\right)
\Biggl\vert_{ren}
=\frac{1}{2\pi i}
\oint_{C}\omega\coth\frac{\hbar\omega}{2k_B T}
d\left[\ln\dpa(\omega,{\kp})
+\ln\dpp(\omega,{\kp})\right],
\label{11}
\eeq
\ni
where the integration path $C$ in the plane of complex $\omega$
is shown in Fig.~1, and the normalized quantities $\Delta_{\|,\bot}$,
having zero values on eigenfrequencies, were substituted defined
by Eqs.(\ref{3})--(\ref{5}), with $\xi_l$ 
changed by $-i\omega$. Note that the function
$\omega\coth[\hbar\omega/(2k_B T)]$ has poles at the imaginary
frequencies $\omega_l=i\xi_l$, $l=\pm 1,\pm 2,\ldots\,$, 
where $\xi_l$ are the Matsubara frequencies (it is, however,
regular at $\omega_0=0$). Because of this, the integration along the
imaginary axis involves semicircles about these poles.
Integration along a semicircle, whose radius extends to infinity,
makes zero contribution to the right-hand side of Eq.(\ref{11}).
As a result Eq.(\ref{11}) leads to
\bes
&&
\sum_{n}\left(\omega_{{\kp},n}^{\|}\coth
\frac{\hbar \omega_{{\kp},n}^{\|}}{2k_B T}
+
\omega_{{\kp},n}^{\bot}\coth
\frac{\hbar \omega_{{\kp},n}^{\bot}}{2k_B T}\right)
\Biggl\vert_{ren}
=\frac{i}{2\pi}
\int_{-\infty}^{\infty}\!\!\xi\cot\frac{\hbar\xi}{2k_B T}
d\left[\ln\dpa(\xi,{\kp})+\ln\dpp(\xi,{\kp})\right]
\nn \\
&&\nn \\
&&\phantom{aaa}
-\sum_{l=1}^{\infty}\mbox{Res}\left[
\omega\coth
\frac{\hbar \omega}{2k_B T}\,
\frac{{\Delta^{\prime}}_{\|}(\omega,{\kp})}{{\dpa}(\omega,{\kp})},
i\xi_l\right]
-\sum_{l=1}^{\infty}\mbox{Res}\left[
\omega\coth
\frac{\hbar \omega}{2k_B T}\,
\frac{{\Delta^{\prime}}_{\bot}(\omega,{\kp})}{{\dpp}(\omega,{\kp})},
i\xi_l\right]
\label{12} \\
&&\nn \\
&&\phantom{aaa}
=-\frac{2k_B T}{\hbar}
\sum_{l=1}^{\infty}\xi_l\left[\frac{1}{{\dpa}(\xi_l,{\kp})}
\frac{\partial {\dpa}(\xi_l,{\kp})}{\partial\xi_l}
+
\frac{1}{{\dpa}(\xi_l,{\kp})}
\frac{\partial {\dpa}(\xi_l,{\kp})}{\partial\xi_l}\right].
\nn
\ees
\ni
Here prime denotes the derivative with respect to $\omega$ and
we took into account that $\Delta_{\|,\bot}$ are even functions
of $\omega$, so that their derivatives are odd ones. This property
leads also to the zero value of the seemingly pure imaginary
integral in the right-hand side of Eq.~(\ref{12}).

Substituting the right-hand side of Eq.(\ref{12}) into Eq.(\ref{10}) 
instead of a nonrenormalized sum, we obtain the renormalized
energy at a temperature $T$ coinciding with Eq.(\ref{8}) derived from
the thermodynamical definition (\ref{6}). In such a manner we
have proved that at certain conditions the thermodynamical energy at
equilibrium is given by the additive sum of the contributions from
the zero-point fluctuations and Planck's photons. The renormalization
of both quantities reduces to the subtraction of the contribution
of a free space with no plates. 
As a consequence, in the absence of dissipation,
photons between plates can be considered as some kind of quasiparticle
excitations in the system of the electromagnetic field interacting
with the matter of plates. The simple example of this situation is
given by metals described by the plasma model.

\section{Energy and free energy for the thermal Casimir force in
the framework of the plasma model}

The considerations of the previous section can be illustrated by the
dielectric function of the plasma model
\beq
{\e}(\omega)=1-\frac{\omega_p^2}{\omega^2},
\qquad
{\e}(i\xi)=1+\frac{\omega_p^2}{\xi^2},
\label{13}
\eeq
\ni
where $\omega_p$ is the plasma frequency. This dielectric function is real
and its parameter does not depend on temperature. The use of the plasma
model to calculate the thermal Casimir force corresponds to the
approach d) described in the introduction. The free electron plasma model
works well for frequencies of visible light and infrared optics.
It is common knowledge that the dominant contribution to the Casimir
effect comes from the range around the characteristic
frequency $\omega_c=c/(2a)$. Thus the plasma model
is applicable in the $a$-range from a few tens of nanometers to
around a hundred micrometers.

For the sake of convenience, we introduce the dimensionless variables
\beq
\tilde\xi=\frac{2a\xi}{c},
\qquad
y^2=4a^2\left({\kp}^2+\frac{\xi^2}{c^2}\right)
\label{14}
\eeq
\ni
in terms of which the plasma dielectric function takes the form
\beq
{\e}(i\tilde\xi)=1+\frac{\tilde{\omega}_p^2}{\tilde{\xi}^2},
\qquad
\tilde{\omega}_p=\frac{2a\omega_p}{c}.
\label{15}
\eeq

In terms of the variables $\tilde\xi,\,y$ the Lifshitz 
formula (\ref{1}) can
be rewritten as
\bes
&&
F_E(a,T)=\frac{k_B T}{16\pi a^2}
\sum_{l=-\infty}^{\infty}
\int_{|{\tilde\xi}_l|}^{\infty}
{y}d{y}\left[ \ln{\dpa}({\tilde\xi}_l,{y})\right.
\label{16} \\
&&
\phantom{aaaaaaaaaaaaaaaaaaaa}\left.
+
\ln{\dpp}({\tilde\xi}_l,{y})\right],
\nn
\ees
\ni
where
\beq
\Delta_{\|,\bot}({\tilde\xi}_l,{y})=1-
r_{\|,\bot}^2({\tilde\xi}_l,{y})e^{-y}
\label{17}
\eeq
\ni
and the reflection coefficients are
\beq
r_{\|}^2({\tilde\xi}_l,{y})=
\left\{\frac{y{\e}(i\tilde\xi)-\sqrt{\left[{\e}(i\tilde\xi)-1\right]
{\tilde\xi}^2
+y^2}}{y{\e}(i\tilde\xi)+\sqrt{\left[{\e}(i\tilde\xi)-1\right]
{\tilde\xi}^2
+y^2}}\right\}^2,
\quad
r_{\bot}^2({\tilde\xi}_l,{y})=
\left\{\frac{y-\sqrt{\left[{\e}(i\tilde\xi)-1\right]{\tilde\xi}^2
+y^2}}{y+\sqrt{\left[{\e}(i\tilde\xi)-1\right]{\tilde\xi}^2
+y^2}}\right\}^2.
\label{18}
\eeq

By virtue of the fact that $\e$ depends on temperature through the
Matsubara frequencies only, the zero-frequency term of Eq.~(\ref{16})
($l=0$) does not contribute into the energy (\ref{6}) [compare
with Eq.~(\ref{7})]. It is notable also that in the special case
of the plasma model the perpendicular reflection coefficient from
Eq.~(\ref{18}) is given by
\beq
r_{\bot}^2({\tilde\xi}_l,{y})=r_{\bot}^2({y})=
\left(\frac{y-\sqrt{{\tilde\omega}_p^2+y^2}}{y+
\sqrt{{\tilde\omega}_p^2+y^2}}\right)^2,
\label{19}
\eeq
\ni
i.e. it is frequency- and temperature-independent for any $l$.
By this reason its derivative with respect to temperature does not
contribute to energy (\ref{6}). As  a result, in the framework of
the plasma model the energy per unit area at a temperature $T$, calculated
by Eqs.~(\ref{6}), (\ref{16}), takes the form
\beq
E^{pl}(a,T)=\frac{k_B T}{8\pi a^2}
\sum_{l=1}^{\infty}
{\tilde\xi}_l^2
\left[ \ln{\dpa}({\tilde\xi}_l,{\tilde\xi}_l)+
\ln{\dpp}({\tilde\xi}_l,{\tilde\xi}_l)\right]
+\frac{k_B T}{4\pi a^2}
\sum_{l=1}^{\infty}
\int_{{\tilde\xi}_l}^{\infty}
{y}d{y}\frac{r_{\|}({\tilde\xi}_l,{y})}{e^y-r_{\|}^2({\tilde\xi}_l,{y})}
\frac{\partial r_{\|}({\tilde\xi}_l,{y})}{\partial{\tilde\xi}_l}.
\label{20}
\eeq
\ni
This equation is convenient for numerical calculations.

Let us now compare the values of energy at temperature $T$ given
by Eq.~(\ref{20}) and free energy of Eq.~(\ref{16}) with the values
of energy at zero temperature given by \cite{2,6,34a,35}
\beq
E(a,0)=\frac{\hbar c}{32\pi^2 a^3}
\int_{0}^{\infty}d{\tilde\xi}
\int_{{\tilde\xi}}^{\infty}
{y}d{y}\left[ \ln{\dpa}({\tilde\xi},{y})
+
\ln{\dpp}({\tilde\xi},{y})\right].
\label{21}
\eeq
\ni
The calculational results at $T=300\,$K for the case of $Al$ with \cite{38}
\beq
\omega_p=11.5\,\mbox{eV}=1.75\times 10^{16}\,\mbox{rad/s}
\label{22}
\eeq
\ni
are shown in Fig.~2. In this figure the dimensionless ratios
\beq
R^{pl}=\frac{E^{pl}(a,T)}{|E^{pl}(a,0)|},{\quad}
\frac{F_E^{pl}(a,T)}{|E^{pl}(a,0)|},{\quad}
\frac{E^{pl}(a,0)}{|E^{pl}(a,0)|}=-1
\label{23}
\eeq
\ni
are plotted by the solid lines 1,\,2 and dashed line, respectively,
as the functions of the surface separation.
The energy at zero temperature $E^{pl}(a,0)$ is computed by Eq.~(\ref{21}) 
where the plasma dielectric function given by Eq.~(\ref{13}) is
substituted. It is clearly seen, that at smallest
separations all three quantities (energy at $T=0$, energy and free
energy at $T=300\,$K) have approximately equal values. With an increase
of the separation distance the modulus of the relative energy at temperature
$T$ decreases to zero limiting value while the modulus of the relative
free energy increases. Note that the limiting cases of small and large
separations can be simultaneously considered as the limits of low and
high temperatures, respectively, if one compares with the so called
effective temperature 
$k_B T_{eff}=\hbar\omega_c=\hbar c/(2a)$ \cite{3,6,30}.

The asymptotic behavior of energy and free energy at small and large
separations (low and high temperatures) in the case of the plasma model
can also be investigated analytically. As was proved in Ref.\cite{39}, 
one can expand Eq.~(\ref{16}) in powers of 
a small parameter $\lambda_p/2\pi a$, where
$\lambda_p$ is the plasma wavelength, and in a contribution, depending
on temperature, it would suffice to preserve the first power only. 
The result valid for all $a\geq\lambda_p$ is
\bes
&&
F_E^{pl}(a,T)=E^{pl}(a,0)-\frac{\hbar c}{8\pi^2a^3}
\sum_{l=1}^{\infty}
\left\{
\vphantom{\left[\frac{2\pi^3}{lt}
\frac{\coth(\pi lt)}{\sinh^2(\pi lt)}\right]}
\frac{\pi}{2(lt)^3}\coth(\pi lt)-\frac{1}{(lt)^4}+
\frac{\pi^2}{2(lt)^2}\frac{1}{\sinh^2(\pi lt)}\right.
\nn \\
&&
\phantom{aa}\left.
+\frac{\lambda_p}{2\pi a}\left[
\frac{\pi}{(lt)^3}\coth(\pi lt)-\frac{4}{(lt)^4}+
\frac{\pi^2}{(lt)^2}\frac{1}{\sinh^2(\pi lt)}+
\frac{2\pi^3}{lt}\frac{\coth(\pi lt)}{\sinh^2(\pi lt)}\right]
\right\},
\label{24}
\ees
\ni
where $t\equiv T_{eff}/T$. The quantity $E^{pl}(a,0)$ is the energy
at zero temperature. Its expansion in powers of $\lambda_p/2\pi a$
can be found in Refs.\cite{6,40} (here the result up to fourth
order should be used in order to get sufficient accuracy at
smallest separations).

From Eq.~(\ref{24}) the required asymptotics follow. At small
separations ($T\ll T_{eff}$) one obtains
\beq
F_E^{pl}(a,T)=E^{pl}(a,0)-\frac{\hbar c\zeta(3)}{16\pi a^3}
\left[\left(1+\frac{\lambda_p}{\pi a}\right)
\left(\frac{T}{T_{eff}}\right)^3
-\frac{\pi^3}{45\zeta(3)}
\left(1+2\frac{\lambda_p}{\pi a}\right)
\left(\frac{T}{T_{eff}}\right)^4\right],
\label{25}
\eeq
\ni
where $\zeta(z)$ is the Riemann zeta function.

Applying the thermodynamical definition (\ref{6}) to Eq.~(\ref{25}),
we obtain the low-temperature asymptotic of energy
\beq
E^{pl}(a,T)=E^{pl}(a,0)+\frac{\hbar c\zeta(3)}{8\pi a^3}
\left[\left(1+\frac{\lambda_p}{\pi a}\right)
\left(\frac{T}{T_{eff}}\right)^3
-\frac{\pi^3}{30\zeta(3)}
\left(1+2\frac{\lambda_p}{\pi a}\right)
\left(\frac{T}{T_{eff}}\right)^4\right].
\label{26}
\eeq
\ni
This asymptotic expression is obtained also from Eqs.~(\ref{8}) or 
(\ref{20}) by the use of the Abel-Plana formula (see Ref.\cite{27}
where similar calculations were performed).

In the opposite case of large separations ($T\gg T_{eff}$),
Eq.~(\ref{24}) leads to the main contribution of the form
\beq
F_E^{pl}(a,T)=-\frac{k_B T}{8\pi a^2}\zeta(3)\left(1-
\frac{\lambda_p}{\pi a}\right).
\label{27}
\eeq
\ni
By virtue of Eq.~(\ref{6}) the asymptotic value of energy is 
$E^{pl}(a,T)=0$. If one wished to have a more exact asymptotic
of energy, the next (exponentially small in $T/T_{eff}$) terms
omitted in Eq.~(\ref{27}) should be taken into account or
Eq.~(\ref{20}) should be used. In both cases the result is one and 
the same
\beq
E^{pl}(a,T)=-k_B T\frac{\pi}{a^2}\left(\frac{T}{T_{eff}}\right)^2
\left(1-2\frac{\lambda_p}{\pi a}\frac{T}{T_{eff}}\right)
e^{-2\pi T/T_{eff}}.
\label{28}
\eeq

Comparison of the numerical calculations presented in Fig.~2 with 
calculations
by the asymptotic formulas of Eqs.~(\ref{25})--(\ref{28}) shows
that the asymptotic of small separations works well within the
separation range $\lambda_p\leq a\leq 2-3\,\mu$m, and the asymptotic
of large separations is applicable for $a\geq 5\,\mu$m.
In the transition range, Eqs.~(\ref{16}), (\ref{20}) should be used
to calculate the values of the free energy and energy for the thermal
Casimir force in the framework of the plasma model.

If we consider the limit $\omega_P\to\infty$ ($\lambda_P\to 0$) in
Eqs.~(\ref{24})--(\ref{28}), the results for ideal metal are obtained.

\section{Different approaches to the calculation of energy and free energy
in the framework of the Drude model}

Let us now consider metals described by the Drude dielectric function
\beq
{\e}(\omega)=1-\frac{\omega_p^2}{\omega(\omega+i\gamma)},
\qquad
{\e}(i\xi)=1+\frac{\omega_p^2}{\xi(\xi+\gamma)},
\label{29}
\eeq
\ni
where $\gamma$ is the relaxation parameter. In terms of a dimensionless
frequency introduced in Eq.~(\ref{14}), the Drude dielectric function
along the imaginary axis is
\beq
{\e}(i\tilde\xi)=1+
\frac{{\tilde\omega}_p^2}{\tilde\xi(\tilde\xi+\tilde\gamma)},
\qquad
\tilde\gamma\equiv\frac{2a}{c}\gamma.
\label{30}
\eeq

As was noticed in the introduction, there is no agreement in 
the recent literature
regarding the use of the Drude model in the framework of the Lifshitz theory.
Because of this, it is appropriate to reexamine the applicability of the
Drude model in the context of the thermal Casimir force. The Drude model,
as opposed to the plasma model, takes into account the phenomenon of
volume relaxation. In reality, this phenomenon plays a role in the 
domain of the normal skin effect where the mean free path $l$ of the
electron is much less than the penetration depth of the electromagnetic
oscillations into a metal $\delta$ and the mean distance $v/\omega$
traveled by an electron in a time $1/2\pi$ of the period of the
electromagnetic field \cite{41,42}. For most of metals at $T=300\,$K
the domain of the normal skin effect extends from the quasistatic fields
(where $\e$ is pure imaginary) to the frequencies of order 
$10^{12}\,$rad/s. What this means is that the Drude dielectric function
has a direct relationship only to plate separations 
$0.1\,\mbox{cm}<a<1\,$km such that the characteristic frequency
$\omega_c=c/(2a)$ belongs to this domain. 
However, at so large separations the Casimir force is extremely small and 
is of academic interest only.

For higher frequencies, depending on which metal is considered, the
anomalous skin effect ($\delta\ll l$, $\delta\ll v/\omega$) or
relaxation region ($v/\omega\ll l\ll\delta$) occur. Here the volume
relaxation described by the parameter $\gamma$ is not significant, but,
in general, the space dispersion gives an important contribution.
Note that in the domain of the anomalous skin effect (it extends up to
around $7\times 10^{13}\,$rad/s) metal cannot be described 
by either the Drude model, given by Eqs.~(\ref{29}), (\ref{30}), or by any
dielectric function depending only on frequency as well.

On further increase of frequency, the transition to the infrared optics
occur, where $v/\omega\ll\delta\ll l$ (or to the ``extremely anomalous
skin effect'' if we use Casimir's terminology \cite{41}).
In this domain the volume relaxation does not play any role. In the 
semiclassical theory of AC conductivity $\e$ is practically real,
signifying no dissipation of the electromagnetic energy within the
metal \cite{43}. Because of this, the plasma model is realistic if
the characteristic frequency $\omega_c$ belongs to the
domain of the infrared optics (see Sec.~III). This domain extends to the
frequencies of around $2\times 10^{16}\,$rad/s and for higher frequencies 
is followed by the domain of the ultraviolet transparency of metals.
However, some interelectron collisions and a scattering on the surface lead
to a small imaginary part of $\e$ in the domain of infrared optics
\cite{44} as is demonstrated by the optical tabulated data for complex
refraction index \cite{38}. These data are often used to find the values
of $\e(i\xi)$ along the imaginary axis through the dispersion relation
\cite{6,22,23,23a,24}. By way of example, for $Al$ the optical data
for $\omega\geq 6.08\times 10^{13}\,$rad/s are tabulated \cite{38}.

At the same time, the existence of the anomalous skin effect domain,
where the concept of $\e(\omega)$ is not applicable, is usually ignored,
and the optical tabulated data are theoretically extended into the domain of
lower frequencies by means of the Drude dielectric function \cite{38}.
This is needed to compute the dispersion integral from zero to infinity.
The values of $\e(i\xi)$ obtained in such a manner by means of the
dispersion relation and extended tabulated data are satisfactory up to
$\xi\sim 10^{15}\,$rad/s with $\e(i\xi)$ obtained by the immediate
substitution of the imaginary frequency into the Drude model according
to Eq.~(\ref{29}) with no use of the dispersion relation. This suggests
that the Drude model can be applied for the calculation of the Casimir
force within a micrometer domain $a\geq 0.4\,\mu$m in parallel with the
plasma model. It should be particularly emphasized, however, that the
application of the Drude model in the domain of infrared optics is
physically unjustified  as the volume relaxation is absent in this
domain (below we call into question also the possibility to substitute 
the Drude dielectric function into the zero-frequency term of the
Lifshitz formula).

In contrast to the case considered in Sec.~II, the Drude metals
are dissipative media, described by the complex $\e(\omega)$. At
a given frequency $\e$ depends explicitly on temperature through
the relaxation parameter $\gamma$. Because of this, the energy of the
equilibrium fluctuating electromagnetic field cannot be presented any
more in the simple form of Eq.~(\ref{9}). In accordance with the
thermodynamic equality (\ref{6}), additional terms appear in the
right-hand side of Eq.~(\ref{9}) containing the derivatives
$\partial\omega_{{\kp},n}^{\|,\bot}/\partial T$ \cite{36}.

Substituting Drude dielectric function (\ref{30}) into the Lifshitz
formula for the free energy (\ref{16}) and using the definition 
(\ref{6}) of energy at temperature $T$, one obtains
\bes
&&
E^{D}(a,T)=-\frac{k_B T^2}{16\pi a^2}\frac{\partial}{\partial T}
f_0^{(\rm{a,b,c})}(a,T)+
\frac{k_B T}{8\pi a^2}
\sum_{l=1}^{\infty}
{\tilde\xi}_l^2\left[\ln{\dpa}({\tilde\xi}_l,{\tilde\xi}_l)+
\ln{\dpp}({\tilde\xi}_l,{\tilde\xi}_l)\right]
\label{31} \\
&&\phantom{aa}+
\frac{k_B T}{4\pi a^2}
\sum_{l=1}^{\infty}
\int_{{\tilde\xi}_l}^{\infty}ydy
\left\{\frac{r_{\|}({\tilde\xi}_l,y)}{e^y-r_{\|}^2({\tilde\xi}_l,y)}
\left[{\tilde\xi}_l
\frac{\partial r_{\|}({\tilde\xi}_l,y)}{\partial {\tilde\xi}_l}+
T\frac{\partial r_{\|}({\tilde\xi}_l,y)}{\partial \tilde\gamma}
\frac{\partial\tilde\gamma}{\partial T}\right]
\right.
\nn \\
&& \phantom{aaaaaa}
\left.
+
\frac{r_{\bot}({\tilde\xi}_l,y)}{e^y-r_{\bot}^2({\tilde\xi}_l,y)}
\left[{\tilde\xi}_l
\frac{\partial r_{\bot}({\tilde\xi}_l,y)}{\partial {\tilde\xi}_l}+
T\frac{\partial r_{\bot}({\tilde\xi}_l,y)}{\partial \tilde\gamma}
\frac{\partial\tilde\gamma}{\partial T}\right]\right\}.
\nn
\ees
\ni
Here the zero-frequency term of Eq.~(\ref{16}) is separated because
there is disagreement in recent literature on whether or not
it contributes to the Casimir energy and force. The immediate consequence
of Eqs.~(\ref{6}), (\ref{16}),  (\ref{30}) [approach a) described in
Introduction] is \cite{25}
\beq
f_0^{(\rm{a})}(a,T)=
\int_{0}^{\infty}dy\,y\ln\left(1-e^{-y}\right)=-\zeta(3).
\label{32}
\eeq
\ni
This result is given by the parallel modes only, while the perpendicular
modes do not contribute.

The special modification of the zero-frequency term of Eq.~(\ref{16}) proposed 
in \cite{30} [approach b)] leads to
\beq
f_0^{(\rm{b})}(a,T)=
-\zeta(3)
+\int_{0}^{\infty}dy\,y\ln\left[1-r_{\bot}^2(y,y)e^{-y}\right].
\label{33}
\eeq
\ni
The two contributions in the right-hand side of Eq.~(\ref{33}) are given
by the parallel (perpendicular) modes, respectively.

If for real metals the same prescription is used as for ideal metal
[approach c)], one obtains \cite{28,29}
\beq
f_0^{(\rm{c})}(a,T)=
2\int_{0}^{\infty}dy\,y\ln\left(1-e^{-y}\right)=-2\zeta(3),
\label{34}
\eeq
\ni
where both polarizations lead to equal nonzero contribution.
Evidently, in the framework of the approaches a) and c) the
zero-frequency term is temperature independent and does not contribute
to the energy (\ref{31}). In the framework of the approach b) there is only
a fair contribution due to the dependence of $r_{\bot}(y,y)$ on
$\gamma(T)$ in Eq.~(\ref{33}).

Note that
Eq.~(\ref{31}) is in direct analogy to Eq.~(\ref{20}). The additional terms 
which are present in Eq.~(\ref{31}) take into account the explicit dependence
of the dielectric permittivity on temperature through the relaxation
parameter. This equation is convenient for the numerical calculations.

Before performing the calculations, let us give the approximate expressions
for both free energy and energy which allow one to compare the results
obtained in the framework of the Drude and plasma models. For this purpose we
expand Eq.~(\ref{16}) in powers of a small parameter $\gamma/\omega_p$
preserving the first-order term only (for $Al$ at $T=300\,$K 
$\gamma=0.05\,\mbox{eV}=7.6\times 10^{13}\,$rad/s, so that for lower
$T$ $\gamma/\omega_p\leq0.004$). The coefficient near this term can be 
computed in the zeroth order in a small parameter
$\alpha=\lambda_p/4\pi a=1/{\tilde\omega}_p$. The result is
\bes
&&
F_E^D(a,T)=F_E^{pl}(a,T)+
\frac{k_B T}{16\pi a^2}
\left\{
\vphantom{\int_{0}^{\infty}}
f_0^{(\rm{a,b,c})}(a,T)
+\zeta(3)-
\int_{0}^{\infty}dy\,y\ln\left[1-r_{\bot}^2(y)e^{-y}\right]\right\}
\nn \\
&&\phantom{aa}
+\frac{\gamma}{\omega_p}\frac{k_B T}{4\pi a^2}
\sum_{l=1}^{\infty}
\left[{\tilde\xi}_l
\int_{{\tilde\xi}_l}^{\infty}
\frac{dy}{e^y-1}+
\frac{1}{{\tilde\xi}_l}
\int_{{\tilde\xi}_l}^{\infty}\!\!
\frac{dy\,y^2}{e^y-1}\right],
\label{35}
\ees
\ni
where $F_E^{pl}(a,T)$ is the free energy in the plasma model
given by Eq.~(\ref{24}), and $r_{\bot}(y)$ is defined in Eq.~(\ref{19}).
It is notable that the results of numerical calculations by this formula 
and by Eqs.~(\ref{16}), (\ref{30}) (with different approaches to the
zero-frequency term) coincide with an accuracy of 0.06\% at
$a=0.4\,\mu$m and better than 0.01\% for $a\geq 3\,\mu$m.
Note, as discussed above, the Drude model leads to satisfactory
$\e(i\xi)$ only up to $\xi\sim 10^{15}\,$rad/s and is in strong
disagreement with the optical tabulated data for higher frequencies.
Because of this, at $T=300\,$K it is meaningless to use the Drude
dielectric function at separations $<0.4\,\mu$m. However, even at these
separations, Eq.~(\ref{35}) is correct with an accuracy of 0.07\%.

As evident from Eqs.~(\ref{32})--(\ref{35}), the contribution of the
zero-frequency term into the difference of the free energies
$\Delta F_E=F_E^D-F_E^{pl}$, computed in the framework of the Drude and
plasma models, depends on the approach used:
\bes
&&
\Delta F_E^{0(\rm{a})}=-\frac{k_B T}{16\pi a^2}
\int_{0}^{\infty}\!\!\!
dy\,y\ln\left[1-r_{\bot}^2(y)e^{-y}\right],
\label{36} \\
&&
\Delta F_E^{0(\rm{b})}=\frac{k_B T}{8\pi a^2}
\frac{\gamma(T)}{\omega_p}
\int_{0}^{\infty}\!\frac{y\,dy}{e^y-1}=
\frac{\pi k_B T}{48 a^2}
\frac{\gamma(T)}{\omega_p},
\label{37} \\
&&
\Delta F_E^{0(\rm{c})}=-\frac{k_B T}{16\pi a^2}
\left\{
\zeta(3)
-\int_{0}^{\infty}\!\!\!
dy\,y\ln\left[1-r_{\bot}^2(y)e^{-y}\right]\right\}.
\label{38}
\ees
\ni
In the case of approaches a) and c),
the difference of the free energies contains linearly decreasing with
temperature terms [Eqs.~(\ref{36}), (\ref{38})]. 
In the case of approach b), owing to the relaxation
parameter, $\Delta F_E^0$ falls off more quickly with decreasing
temperature [the same is true for $F_E^{pl}$, as is seen from
Eq.~(\ref{25}), and for the summation term in the right-hand side
of Eq.~(\ref{35})]. It should be particularly emphasized that the 
presence of the linear terms in temperature in the free energy is in 
contradiction with the requirements of thermodynamics (see Sec.~V).

To obtain the approximate perturbative expression for the energy by
analogy with Eq.~(\ref{35}), one should use the explicit dependence of
$\gamma$ on temperature. It has been known that at temperature
$T>T_D/4$, where $T_D$ is the Debye temperature (for $Al$
$T_D=428\,$K \cite{45}), $\partial\gamma/\partial T=\gamma/T$, i.e.
$\gamma$ is linear in temperature. Generally, 
$\partial\gamma/\partial T=\nu\gamma/T$ with $\nu=\nu(T)\geq 1$.
In Fig.~3, the dependence of $\gamma$ on temperature is plotted for $Al$
on the basis of tabulated data \cite{45}. Finally, the required expression
for the energy is
\beq
E^D(a,T)=E^{pl}(a,T)+e_0^{(\rm{a,b,c})}(a,T)
+\frac{k_B T}{4\pi a^2}\frac{\gamma}{\omega_p}
\sum_{l=1}^{\infty}
\left[\frac{2{\tilde\xi}_l^2}{e^{{\tilde\xi}_l}-1}
-
(\nu+1){\tilde\xi}_l
\int_{{\tilde\xi}_l}^{\infty}
\frac{dy}{e^y-1}-
\frac{\nu -1}{{\tilde\xi}_l}
\int_{{\tilde\xi}_l}^{\infty}
\frac{dy\,y^2}{e^y-1}
\vphantom{\frac{2{\tilde\xi}_l^2}{e^{{\tilde\xi}_l}-1}}
\right],
\label{39}
\eeq
\ni
where $e_0^{(\rm{a})}=e_0^{(\rm{c})}=0$, and
\beq
e_0^{(\rm{b})}(a,T)=-\frac{\pi k_B T}{48a^2}\,\frac{\nu\gamma}{\omega_p}.
\label{40}
\eeq

In Fig.~4 the results of the numerical calculations are shown for $Al$
described by the Drude model in different approaches at $T=300\,$K.
In the vertical axis the dimensionless ratios are plotted
\beq
R^D=\frac{E^D(a,T)}{|E^D(a,0)|},{\quad}
\frac{F_E^D(a,T)}{|E^D(a,0)|},{\quad}
\frac{E^D(a,0)}{|E^D(a,0)|}=-1
\label{41}
\eeq
\ni
as a function  of the surface separation.
Curve 1 shows the behavior of the relative energy (which is practically
the same in all three approaches); curves 2a, 2b and 2c show the
relative free energy in the approaches a), b), and c), respectively.
The dashed curve is for the energy at zero temperature. All calculations
are performed both using the exact expressions (\ref{16}), (\ref{31}) and
the approximate ones (\ref{35}), (\ref{39}) with coinciding results.

It is important to explain in more detail the notation $E^D(a,0)$.
It is the value of energy in the framework of the Drude model (\ref{30}), 
computed at zero temperature in the sense that Eq.~(\ref{21}) with
a double integral instead of a discrete sum is employed.
At the same time, in calculations of $E^D(a,0)$ the value of the
relaxation parameter $\gamma$ at $T=300\,$K is used. 
We divide the calculational results into this hybrid quantity, previously
used in literature (see, e.g., Refs.~\cite{6,22,23,23a,24}).
This allows one to associate this quantity with energy and free energy
in order to clarify its physical meaning.

As is seen from Fig.~4, curve 1, illustrating the equal behavior
of energy in all three approaches, and curve 2b, illustrating the
behavior of free energy as given by the approach b), demonstrate
plausible properties. Among other things, the free energy approaches energy
with a decrease of the surface separation distance (compare with Fig.~2
in the case of the plasma model). As to the curves 2a and 2c,
representing the free energy in the approaches a) and c), they do not
approach to each other nor to energy within the application range of the
Drude dielectric function. Note that even at separations of about
$0.4-0.5\,\mu$m the free energy $F_E^D$ (curve 2a), obtained by the
direct application of the Lifshitz formula, differs by 8\% from
the double integral $E^D(a,0)$ (dashed line).

An important point is that not only the free energy of curve 2a
but also 2b and 2c, and energy of curve 1 do not approach the
dashed line in Fig.~4 representing the quantity
which is in common use as a measure of
energy at zero temperature \cite{6,22,23,23a,24}. This is clearly seen from
Fig.~5 where the curves 1, 2b and 2c are reproduced on an enlarged
scale for the smallest separations where the Drude model is applicable.
The long-dashed curve 3 in Fig.~5 illustrates the dependence of one more
quantity on surface separation defined as
\beq
R^D=\frac{E_{\gamma}^D(a,T)}{|E^D(a,0)|},
\label{42}
\eeq
\ni
where $E_{\gamma}^D$ is the energy at a temperature $T$ computed on the
assumption that $\gamma$ does not depend on temperature (and preserves
its value as at $T=300\,$K). Curve 3 is computed by Eq.~(\ref{31}) with
$\partial\tilde\gamma/\partial T=0$. The same curve is obtained by the
application of the approximate Eq.~(\ref{39}) with $\nu=0$.

From Fig.~5 we notice that curve 3 approaches the short-dashed curve with 
a decrease of a separation distance. Because of this, it may be concluded 
that the hybrid quantity $E^D(a,0)$ computed in the literature is in 
fact some approximation for $E_{\gamma}^D$, i.e. for the energy at temperature
$T$ computed without regard for the explicit dependence of the dielectric
properties on temperature [remind that this kind dependence is absent
in the case of the plasma model (see Secs.~II, III) but is essential for
metals described by the Drude model]. From Fig.~5 it follows that at a 
separation of $0.5\,\mu$m $E^D(a,0)$ departs from the correct value of
energy (curve 1) by approximately 0.75\%. As to the free energies of
the approaches a) and c), the deviations are larger (8\% and 3.3\%,
respectively; these approaches are in contradiction with thermodynamics,
see Sec.~V). The above deviations should be added to the errors of
$E^D(a,0)$, discussed in Ref.~\cite{23}, that are connected with 
uncertainties in the optical tabulated data. 

\section{Entropy for the thermal Casimir force between real metals}

Considerations of the entropy of the fluctuating field
in dependence on temperature allows one to
test different approaches discussed above for conformity to
thermodynamics. Entropy of the fluctuating electromagnetic field can be
expressed in terms of a free energy
\beq
S(a,T)=-\frac{\partial F_E(a,T)}{\partial T}
\label{43}
\eeq
\ni
or, taking into account Eq.~(\ref{6}), identically, as
\beq
S(a,T)=-\frac{1}{T}\left[E(a,T)-F_E(a,T)\right].
\label{44}
\eeq
\ni 
So it can be simply computed by the use of the results for the free energy
and energy obtained in Secs.~III, IV. 

Let us start with the plasma model where the analytical calculation
is possible [approach d)]. At small separations (low temperatures) one
can use Eqs.~(\ref{25}), (\ref{26}) for the free energy and energy,
respectively ($a\geq\lambda_p$ is supposed). Then both Eqs.~(\ref{43})
and (\ref{44}) lead to one and the same result
\beq
S^{pl}(a,T)=\frac{3k_B \zeta(3)}{8\pi a^2}
\left(\frac{T}{T_{eff}}\right)^2\left\{1-
\frac{4\pi^3}{135\zeta(3)}\frac{T}{T_{eff}}
+\frac{\lambda_p}{\pi a}
\left[1-\frac{8\pi^3}{135\zeta(3)}\frac{T}{T_{eff}}\right]\right\}.
\label{45}
\eeq
\ni
Note that this expression was first obtained in Ref.~\cite{32}
with errors in numerical coefficients, because in Ref.~\cite{32} 
the energy of thermal photons was not taken properly into account.
At large separations (high temperatures) the asymptotic expressions
(\ref{27}), (\ref{28}) are applicable leading to
\beq
S^{pl}(a,T)=\frac{k_B \zeta(3)}{8\pi a^2}
\left(1-\frac{\lambda_p}{\pi a}\right)
\label{46}
\eeq
\ni
(we have omitted exponentially small terms in $2\pi T/T_{eff}$).
In the limit of $\lambda_p\to 0$ Eqs.~(\ref{45}), (\ref{46}) lead to
the values of entropy for plates made of ideal metal
\beq
S(a,T)=\frac{3k_B \zeta(3)}{8\pi a^2}
\left(\frac{T}{T_{eff}}\right)^2\left[1-
\frac{4\pi^3}{135\zeta(3)}\frac{T}{T_{eff}}\right],
\qquad
S(a,T)=\frac{k_B \zeta(3)}{8\pi a^2}
\label{47}
\eeq
\ni
for $T\ll T_{eff}$, $T\gg T_{eff}$, respectively.
The results (47) coincide with those obtained for an ideal metal
in Ref.~\cite{46}. Asymptotical behavior of the entropy for an ideal
metal in a high temperature limit was obtained also in Ref.~\cite{47}.
The result of \cite{47} is, however, two times smaller than in
Eq.~(\ref{47}) and Ref.~\cite{46} due to an error contained 
not only in the entropy but
also in the expression for the Casimir energy between two plates made of
ideal metal at zero temperature as is used in Ref.~\cite{47}.

It is obvious that Eq.~(\ref{45}) leads to nonnegative values of
entropy with $S^{pl}(a,0)=0$ as is demanded by the third law of 
thermodynamics (the Nernst heat theorem \cite{48}).

We now direct our attention to the entropy in the framework of the Drude
model. As before, numerical calculations can be performed by the 
exact formulas for 
the energy and free energy or by the approximate ones valid at 
$a\geq\lambda_p$ with coinciding results. From Eqs.~(\ref{35}), (\ref{43})
one obtains
\beq
S^D(a,T)=S^{pl}(a,T)+S_0^{(\rm{a,b,c})}(a,T)
+\frac{k_B}{4\pi a^2}\frac{\gamma}{\omega_p}
\sum_{l=1}^{\infty}
\left[\frac{2{\tilde\xi}_l^2}{e^{{\tilde\xi}_l}-1}
-
(\nu+2){\tilde\xi}_l
\int_{{\tilde\xi}_l}^{\infty}
\frac{dy}{e^y-1}-
\frac{\nu}{{\tilde\xi}_l}
\int_{{\tilde\xi}_l}^{\infty}
\frac{dy\,y^2}{e^y-1}
\vphantom{\frac{2{\tilde\xi}_l^2}{e^{{\tilde\xi}_l}-1}}
\right].
\label{48}
\eeq
\ni
Here $S^{pl}(a,T)$ is the entropy in the framework of the plasma model 
computed by Eqs.~(\ref{24}), (\ref{43}), and $S_0^{(\rm{a,b,c})}(a,T)$,
defined by
\beq
S_0^{(\rm{a,b,c})}(a,T)=-\frac{k_B}{16\pi a^2}
\frac{\partial}{\partial T}
\left\{T\left[
\vphantom{\int_{0}^{\infty}}
f_0^{(\rm{a,b,c})}(a,T)
+\zeta(3)-
\int_{0}^{\infty}dy\,y\ln\left(1-r_{\bot}^2(y)e^{-y}\right)
\right]\right\},
\label{49}
\eeq
\ni
describes the contribution of the zero-frequency term of the Lifshitz
formula into entropy in different approaches. Using the same 
perturbation expansions as in Sec.~IV, one obtains
\bes
&&
S_0^{(\rm{a})}(a,T)=-\frac{k_B\zeta(3)}{16\pi a^2}
\left(1-2\frac{\lambda_p}{\pi a}+3\frac{\lambda_p^2}{\pi^2 a^2}\right),
\nn \\
&&
S_0^{(\rm{b})}(a,T)=-\frac{k_B\pi}{48 a^2}(\nu+1)\frac{\gamma}{\omega_p},
\label{50} \\
&&
S_0^{(\rm{c})}(a,T)=\frac{k_B\zeta(3)}{8\pi a^2}\frac{\lambda_p}{\pi a}
\left(1-\frac{3}{2}\frac{\lambda_p}{\pi a}\right).
\nn
\ees

The results of numerical calculations using Eqs.~(\ref{48}), (\ref{50}) 
for $a=2\,\mu$m are 
presented in Fig.~6. As is seen from the figure, in the approach a)
entropy is negative in a wide temperature range from $T=0$ to
almost $T=300\,$K, which is a nonphysical result. In the approach a)
entropy preserves the negative sign for lesser separations between
the plates as well.
In the approaches
b), c) entropy is positive as it must be. In the approach b)
$S^D(a,0)=0$, whereas in the approaches a), c) $S^D(a,0)\neq 0$
which is in contradiction with the Nernst heat theorem. From
Eqs.~(\ref{48}), (\ref{50}) it follows that
\bes
&&
S^D(a,0)=S_0^{(\rm{a,b,c})}(a,0), 
\label{51} \\
&&
S_0^{(\rm{b})}(a,0)=0,\quad
S_0^{(\rm{c})}(a,0)-S_0^{(\rm{a})}(a,0)=\frac{k_B\zeta(3)}{16\pi a^2},
\nn 
\ees
\ni
where the absolute values of $S_0^{(\rm{a,c})}(a,0)$ are given by 
Eq.~(\ref{50}). They are not only different from zero but depend on
the parameters of the system (plate separation distance and plasma
wavelength) which is prohibited by the third law of thermodynamics
\cite{48}.
Because of this, approaches a) and c) must be rejected. Note also
that approach a) predicts nonzero value of entropy at zero temperature
for an ideal metal in contradiction with the field-theoretical result
of Ref.~\cite{46}. As for approaches b) and d), based on the special
modification of the zero-frequency term of the Lifshitz formula and on the
use of the plasma model, respectively, they are in agreement with the
requirements of thermodynamics. To decide between them some additional
considerations, which are presented in the next section,
 are needed.

\section{Conclusions and discussion}

In the above the correlation between the Casimir energy and 
the free energy at a 
temperature $T$ is investigated for the case of two plane parallel plates
made of real metal. It is shown that for the nondissipative media described by
the real dielectric permittivity with no explicit dependence on temperature
the photons between plates can be considered as the elementary excitations
of the electromagnetic field interacting with a matter of plates.
In this case the energy at temperature $T$ is proved to be a sum of the
(renormalized) energy of zero-point oscillations and thermal photons.
If the media are dissipative and their dielectric permittivity depends
on temperature, the simple picture above is not correct. The concept
of thermal photons loses immediate significance and the energy of
fluctuating field contains additional terms depending on the derivatives
of the dielectric permittivity with respect to temperature.

The expression for the energy at a temperature $T$ found in this paper helps to
elucidate the meaning of the so called ``Casimir energy at zero temperature''
calculated by many authors as a double integral using the Drude model and
optical tabulated data at room temperature (note that this quantity is of
great importance as it is proportional to the Casimir force in the
configuration of a sphere or a spherical lens above a plate used in
experiments \cite{6,7,8,9,10,11,12,15}). The commonly accepted opinion that
the above-mentioned quantity is approximately equal to the free energy at 
small temperatures (small separations) is inexact.
In fact, even at rather small separations the ``Casimir energy at zero 
temperature'' deviates from the free energy by several percent but
approaches to the energy at room temperature calculated on the assumption
that the dielectric permittivity does not depend on $T$ explicitly
(this assumption is not correct in the case of the Drude dielectric
function).

Different approaches to describe the thermal Casimir force from recent
literature were compared and analysed [approaches a), b), c) in the
framework of the Drude model and approach d) in the framework of the
plasma model - see Introduction]. The quantitative expressions for the 
entropy of the
fluctuating field are obtained here for the first time in the case
of real metals. They give the possibility to conclude that the approaches 
a) and c)  are in contradiction with the principles of thermodynamics and
must be rejected. The approaches b) and d) are found to be in agreement with
thermodynamics.

To make a choice between the approaches b) and d) let us discuss the
behavior of the dielectric permittivities of the plasma and Drude models at
small frequencies. Several authors \cite{25,28,29,33} give preference 
to the Drude
model because it shows $\omega^{-1}$ frequency dependence of the dielectric
permittivity at small frequencies as it follows from Maxwell equations
(compare with $\omega^{-2}$ frequency dependence given by the plasma model).
Although this statement is true, it should be remembered that the Drude model
is not applicable at all frequencies. We note that the concept of
$\e(\omega)$ itself, not only the Drude model, does not work in the 
domain of the anomalous skin effect (see Sec.~IV). As to the quasistatic
limit, although $\e(\omega)$ is of order $\omega^{-1}$ in this domain,
the Drude model is also not applicable as the correct $\e(\omega)$ is
pure imaginary. Since the zero-frequency term of the Lifshitz formula
necessarily belongs to the domain of the quasistatic fields, where the 
concept of traveling waves fails, the substitution of the Drude dielectric
function into this term (resulting in all the above problems) seems to be
unjustified.

To clarify the situation with the thermal Casimir force, let us consider
two pairs of plane parallel plates $a=5\,\mu$m apart made of $Al$ (one
pair of plates) and of indium tin oxide (the other one).  Due to large
separation distance, the asymptotic of high temperatures is applicable
and only the zero-frequency term of the Lifshitz formula determines 
the total value of the Casimir force. At quasistatic frequencies both $Al$
and indium tin oxide are good conductors. Because of this, the Lifshitz
formula would lead to one and the same Casimir force at $5\,\mu$m
separation for both pairs of plates if one substitutes into it the
actual reflection properties of these materials at zero frequency.
This conclusion is in contradiction with intuition.
Note that an indium tin oxide is transparent to visible and near
infrared light. Within a wide wavelength range
7$\,\mu$m$<\lambda<100\,\mu$m around the characteristic wavelength
$\lambda_c=62.8\,\mu$m (the latter corresponds to the characteristic
frequency $\omega_c=c/2a=3\times 10^{13}\,$rad/s), giving the main 
contribution into the Casimir force at zero temperature \cite{3},
the reflectivity of indium tin oxide is below 80\% \cite{49}.
Note that the second parameter of the problem, first Matsubara frequency,
is $\omega_M=2\pi k_BT/\hbar=2.45\times 10^{14}\,$rad/s, i.e. 
$\lambda_M=7.7\,\mu$m, which belongs to the region of even larger transparency
of indium tin oxide. In this situation it is difficult to imagine that
at $a=5\,\mu$m the indium tin oxide plates are attracted with the same
Casimir force as $Al$ plates which are almost perfect reflectors
within a wide range around the characteristic wavelength.

We can avoid this contradiction between the literally understood
theory and physical intuition if we assume that 
it is not correct to substitute the actual behavior of
the dielectric permittivity at zero frequency into
the Lifshitz formula. Instead, in order to obtain the physically
correct results, the frequency 
dependence of the dielectric permittivity
and reflection coefficients around the
characteristic frequency should be extrapolated to zero Matsubara
frequency and substituted into the Lifshitz formula.
If this conjecture is accepted, one should conclude that within the 
range of micrometer separation distances between plates the plasma
model dielectric function, i.e. the approach d), is preferable as
compared with the use of the Drude dielectric function combined with
any of the above approaches a), b), c). It is apparent from the fact that 
the plasma dielectric function and respective reflection coefficients admit
reasonable continuation from the range of infrared optics to zero
frequency.

The contradictions discussed in this paper lead to a conclusion that the 
concepts of the frequency dependent dielectric permittivity 
and fluctuating electromagnetic field inside media in application
to the thermal Casimir force between real metals are inadequate idealizations. 
Less sophisticated approaches, such as the surface impedance approach 
(the Leontovich boundary conditions), which does not consider the fluctuating
field inside matter \cite{3,31}, appear to be more adequate and lead to
physically justified results for all separation distances between plates.
By way of example, in the domain of the infrared optics the surface 
impedance leads to the same results as the Lifshitz formula in combination 
with the plasma model [approach d)]. If the characteristic frequency
belongs to the domain of the normal skin effect, where the Drude model is
physically correct, there is no reasonable continuation of $\e$ to zero
frequency avoiding the above problems. At the same time, the impedance
approach, when applied in the domain of the normal skin effect, leads to
quite satisfactory results \cite{31} coinciding with those for ideal
metal as it must be at separations larger than 0.1\,cm [almost the same
results are given in this domain by the approaches b), c)].

To conclude, at present the impedance approach can be considered as the 
most universal, reliable and straightforward way to calculate the thermal
Casimir force between real metals at different separation distances.
In the domain of micrometer separations the plasma model is also
realistic. Regarding the Drude model, it can be used to describe the thermal
Casimir force only with some appropriate modification of the 
zero-frequency term of the Lifshitz formula [like in the approach b), for 
instance].

\section*{ACKNOWLEDGMENTS}
 G.L.K.~is greatly indebted to I.~E.~Dzyaloshinskii for helpful discussions. 
The authors are grateful to U.~Mohideen for valuable remarks.
They acknowledge the financial support from CNPq.


\newpage
\widetext
\begin{figure}[h]
\vspace*{-7cm}
\epsfxsize=20cm\centerline{\epsffile{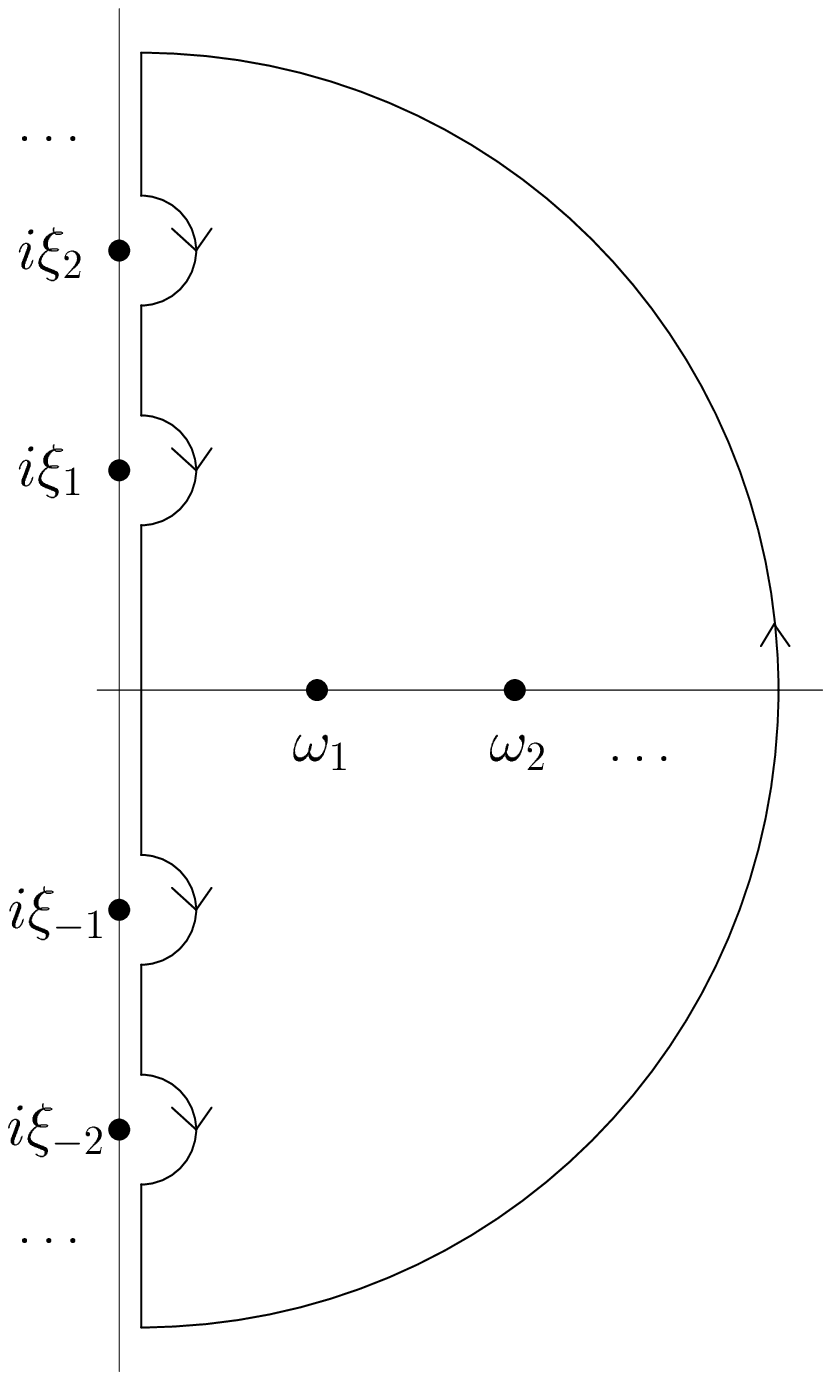}}
\vspace*{-5cm}
\caption{Integration path $C$ in the plane of complex frequency.
The Matsubara frequencies are $\xi_l$ and photon eigenfrequencies
are $\omega_n$.
}
\end{figure}
\newpage
\widetext
\begin{figure}[h]
\vspace*{-7cm}
\epsfxsize=20cm\centerline{\epsffile{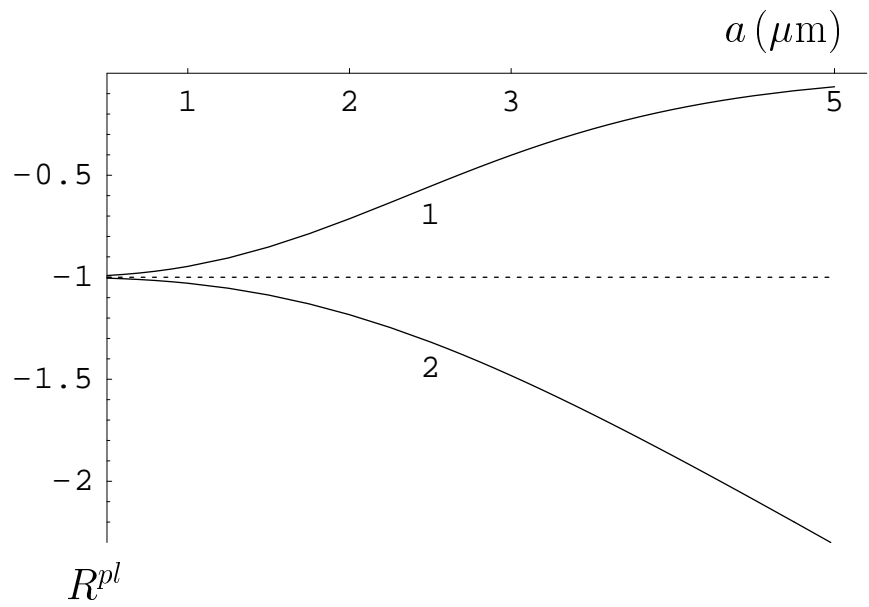}}
\vspace*{-8cm}
\caption{Relative energy at temperature $T=300\,$K (curve 1),
free energy (curve 2), and energy at zero temperature (dashed line) 
versus surface separation in the framework of the plasma model.
}
\end{figure}
\newpage
\widetext
\begin{figure}[h]
\vspace*{-7cm}
\epsfxsize=20cm\centerline{\epsffile{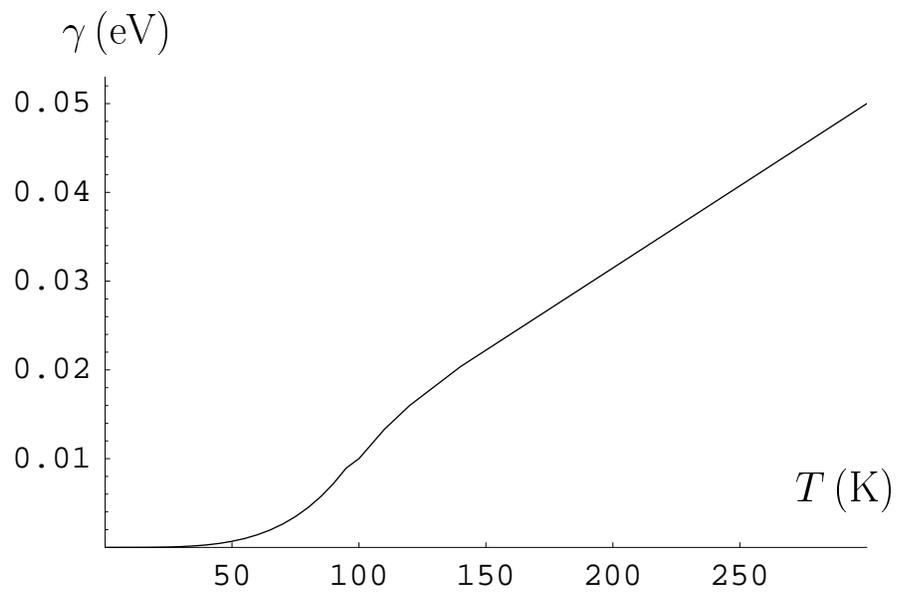}}
\vspace*{-8cm}
\caption{Relaxation parameter of $Al$  
versus temperature.
}
\end{figure}
\newpage
\widetext
\begin{figure}[h]
\vspace*{-7cm}
\epsfxsize=20cm\centerline{\epsffile{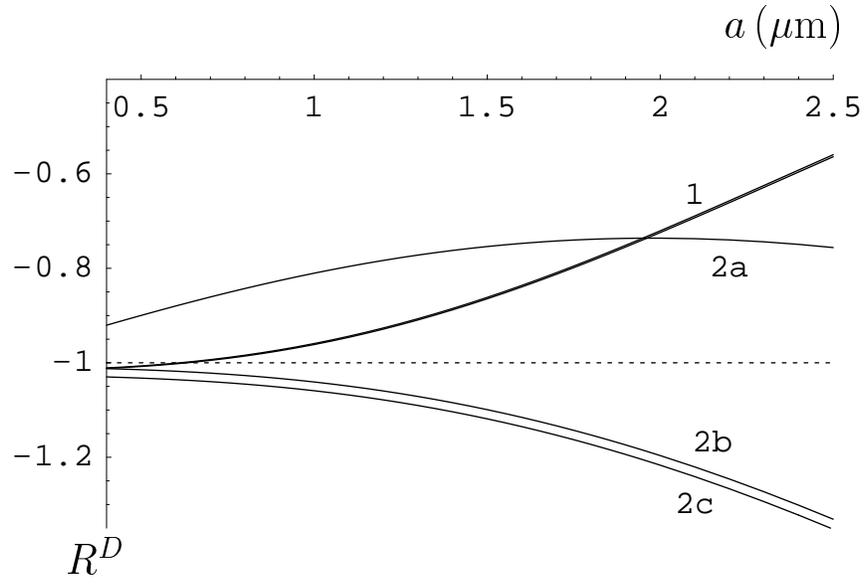}}
\vspace*{-8cm}
\caption{Relative energy at temperature $T=300\,$K (curve 1),
free energy [curve 2a in the approach a), curve 2b in the approach b),
and curve 2c in the approach c)], and 
energy at zero temperature (dashed line) 
versus surface separation in the framework of the Drude model.
}
\end{figure}
\newpage
\widetext
\begin{figure}[h]
\vspace*{-7cm}
\epsfxsize=20cm\centerline{\epsffile{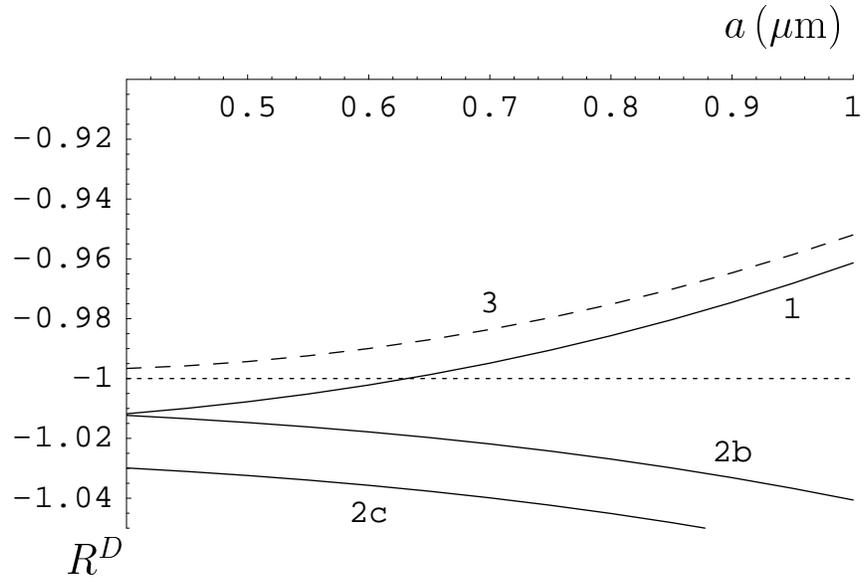}}
\vspace*{-8cm}
\caption{Relative energy at temperature $T=300\,$K (curve 1),
free energy [curve 2b in the approach b)
and curve 2c in the approach c)], and 
energy at zero temperature (short-dashed line) 
versus surface separation in the framework of the Drude model
reproduced on an enlarged scale.
Long-dashed curve 3 presents energy computed on the assumption
that dielectric permittivity does not depend on temperature.
}
\end{figure}
\newpage
\widetext
\begin{figure}[h]
\vspace*{-7cm}
\epsfxsize=20cm\centerline{\epsffile{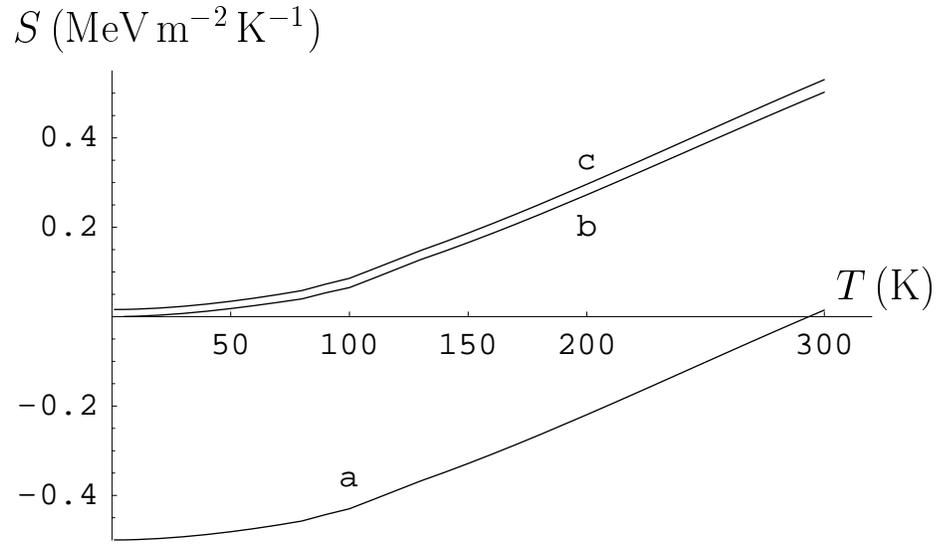}}
\vspace*{-8cm}
\caption{Entropy of fluctuating electromagnetic field in the 
framework of the Drude model versus temperature computed on the basis
of approaches a), b), and c) (curves a, b and c, respectively).
}
\end{figure}
\end{document}